\documentclass[ALICE,manyauthors]{cernphprep}
\usepackage[comma,square,numbers,sort&compress]{natbib}
\usepackage{hyperref}
\usepackage{lineno}
\usepackage{xspace}
\usepackage[T1]{fontenc}
\usepackage{orcidlink}
\begin{document}
%

\newcommand{\pp}           {pp\xspace}
\newcommand{\ppbar}        {\mbox{$\mathrm {p\overline{p}}$}\xspace}
\newcommand{\XeXe}         {\mbox{Xe--Xe}\xspace}
\newcommand{\PbPb}         {\mbox{Pb--Pb}\xspace}
\newcommand{\pA}           {\mbox{pA}\xspace}
\newcommand{\pPb}          {\mbox{p--Pb}\xspace}
\newcommand{\AuAu}         {\mbox{Au--Au}\xspace}
\newcommand{\dAu}          {\mbox{d--Au}\xspace}

\newcommand{\s}            {\ensuremath{\sqrt{s}}\xspace}
\newcommand{\snn}          {\ensuremath{\sqrt{s_{\mathrm{NN}}}}\xspace}
\newcommand{\pt}           {\ensuremath{p_{\rm T}}\xspace}
\newcommand{\meanpt}       {$\langle p_{\mathrm{T}}\rangle$\xspace}
\newcommand{\ycms}         {\ensuremath{y_{\rm CMS}}\xspace}
\newcommand{\ylab}         {\ensuremath{y_{\rm lab}}\xspace}
\newcommand{\etarange}[1]  {\mbox{$\left | \eta \right |~<~#1$}}
\newcommand{\yrange}[1]    {\mbox{$\left | y \right |~<~#1$}}
\newcommand{\dndy}         {\ensuremath{\mathrm{d}N_\mathrm{ch}/\mathrm{d}y}\xspace}
\newcommand{\dndeta}       {\ensuremath{\mathrm{d}N_\mathrm{ch}/\mathrm{d}\eta}\xspace}
\newcommand{\avdndeta}     {\ensuremath{\langle\dndeta\rangle}\xspace}
\newcommand{\dNdy}         {\ensuremath{\mathrm{d}N_\mathrm{ch}/\mathrm{d}y}\xspace}
\newcommand{\Npart}        {\ensuremath{N_\mathrm{part}}\xspace}
\newcommand{\Ncoll}        {\ensuremath{N_\mathrm{coll}}\xspace}
\newcommand{\dEdx}         {\ensuremath{\textrm{d}E/\textrm{d}x}\xspace}
\newcommand{\RpPb}         {\ensuremath{R_{\rm pPb}}\xspace}

\newcommand{\nineH}        {$\sqrt{s}~=~0.9$~Te\kern-.1emV\xspace}
\newcommand{\seven}        {$\sqrt{s}~=~7$~Te\kern-.1emV\xspace}
\newcommand{\twoH}         {$\sqrt{s}~=~0.2$~Te\kern-.1emV\xspace}
\newcommand{\twosevensix}  {$\sqrt{s}~=~2.76$~Te\kern-.1emV\xspace}
\newcommand{\five}         {$\sqrt{s}~=~5.02$~Te\kern-.1emV\xspace}
\newcommand{\twosevensixnn}{$\sqrt{s_{\mathrm{NN}}}~=~2.76$~Te\kern-.1emV\xspace}
\newcommand{\fivenn}       {$\sqrt{s_{\mathrm{NN}}}~=~5.02$~Te\kern-.1emV\xspace}
\newcommand{\LT}           {L{\'e}vy-Tsallis\xspace}
\newcommand{\GeVc}         {Ge\kern-.1emV/$c$\xspace}
\newcommand{\MeVc}         {Me\kern-.1emV/$c$\xspace}
\newcommand{\TeV}          {Te\kern-.1emV\xspace}
\newcommand{\GeV}          {Ge\kern-.1emV\xspace}
\newcommand{\MeV}          {Me\kern-.1emV\xspace}
\newcommand{\GeVmass}      {Ge\kern-.2emV/$c^2$\xspace}
\newcommand{\MeVmass}      {Me\kern-.2emV/$c^2$\xspace}
\newcommand{\lumi}         {\ensuremath{\mathcal{L}}\xspace}

\newcommand{\ITS}          {\rm{ITS}\xspace}
\newcommand{\TOF}          {\rm{TOF}\xspace}
\newcommand{\ZDC}          {\rm{ZDC}\xspace}
\newcommand{\ZDCs}         {\rm{ZDCs}\xspace}
\newcommand{\ZNA}          {\rm{ZNA}\xspace}
\newcommand{\ZNC}          {\rm{ZNC}\xspace}
\newcommand{\SPD}          {\rm{SPD}\xspace}
\newcommand{\SDD}          {\rm{SDD}\xspace}
\newcommand{\SSD}          {\rm{SSD}\xspace}
\newcommand{\TPC}          {\rm{TPC}\xspace}
\newcommand{\TRD}          {\rm{TRD}\xspace}
\newcommand{\VZERO}        {\rm{V0}\xspace}
\newcommand{\VZEROA}       {\rm{V0A}\xspace}
\newcommand{\VZEROC}       {\rm{V0C}\xspace}
\newcommand{\Vdecay} 	   {\ensuremath{V^{0}}\xspace}

\newcommand{\hyp}         {${}^{3}_{\Lambda}\mathrm{H}\xspace}
\newcommand{\ee}           {\ensuremath{e^{+}e^{-}}} 
\newcommand{\pip}          {\ensuremath{\pi^{+}}\xspace}
\newcommand{\pim}          {\ensuremath{\pi^{-}}\xspace}
\newcommand{\kap}          {\ensuremath{\rm{K}^{+}}\xspace}
\newcommand{\kam}          {\ensuremath{\rm{K}^{-}}\xspace}
\newcommand{\pbar}         {\ensuremath{\rm\overline{p}}\xspace}
\newcommand{\kzero}        {\ensuremath{{\rm K}^{0}_{\rm{S}}}\xspace}
\newcommand{\lmb}          {\ensuremath{\Lambda}\xspace}
\newcommand{\almb}         {\ensuremath{\overline{\Lambda}}\xspace}
\newcommand{\Om}           {\ensuremath{\Omega^-}\xspace}
\newcommand{\Mo}           {\ensuremath{\overline{\Omega}^+}\xspace}
\newcommand{\X}            {\ensuremath{\Xi^-}\xspace}
\newcommand{\Ix}           {\ensuremath{\overline{\Xi}^+}\xspace}
\newcommand{\Xis}          {\ensuremath{\Xi^{\pm}}\xspace}
\newcommand{\Oms}          {\ensuremath{\Omega^{\pm}}\xspace}
\newcommand{\degree}       {\ensuremath{^{\rm o}}\xspace}

\begin{titlepage}
\PHyear{2023}       
\PHnumber{020}      
\PHdate{17 February}  

\title{Measurement of the $\Lambda$ hyperon lifetime}
\ShortTitle{Measurement of the $\Lambda$ hyperon lifetime}   

\Collaboration{ALICE Collaboration\thanks{See Appendix~\ref{app:collab} for the list of collaboration members}}
\ShortAuthor{ALICE Collaboration} 

\begin{abstract}
A new, more precise measurement of the $\Lambda$ hyperon lifetime is performed using a large data sample of Pb--Pb collisions at $\sqrt{s_{\rm NN}}$ $=$ 5.02 TeV with ALICE. The $\Lambda$ and $\overline{\Lambda}$ hyperons are reconstructed at midrapidity using their two-body weak decay channel $\Lambda \rightarrow \mathrm{p} + \pi^{-}$ and $\overline{\Lambda} \rightarrow \overline{\mathrm{p}} + \pi^{+}$. The measured value of the $\Lambda$ lifetime is $\tau_{\Lambda} = [261.07 \pm   0.37 \ ( \rm stat.) \pm 0.72 \ (\rm syst.) ]\ \rm ps$. The relative difference between the lifetime of $\Lambda$ and $\overline{\Lambda}$, which represents an important test of CPT invariance in the strangeness sector, is also measured. The obtained value $(\tau_{\Lambda}-\tau_{\overline{\Lambda}})/\tau_{\Lambda} = 0.0013 \pm 0.0028 \ (\mathrm{stat.})  \pm 0.0021 \ (\mathrm{syst.})$ is consistent with zero within the uncertainties. Both measurements 
of the $\Lambda$ hyperon lifetime and of the relative difference between $\tau_{\Lambda}$ and $\tau_{\overline{\Lambda}}$ are in agreement with the corresponding world averages of the Particle Data Group and about a factor of three more precise.

\end{abstract}
\end{titlepage}

\setcounter{page}{2}  

\section{Introduction}
\label{sec:introduction}

The $\Lambda$ is the lightest hyperon, with strangeness $S=-1$, isospin $I=0$, and quark content $\mathrm{uds}$. Its lifetime has been measured in past experiments starting from 1963 using its weak decay channels $\Lambda \rightarrow \mathrm{p}  + \pi^{-}$ and $\overline{\Lambda} \rightarrow \overline{\mathrm{p}} + \pi^{+}$. The world average reported in the Review of Particle Physics of the Particle Data Group (PDG)~\cite{PDG} is $\tau_{\Lambda} = 263.2 \pm 2.0$ ps. This is the result of averaging the measurements performed in 1973 by Poulard et al.~\cite{Poulard} and 1975 by Clayton et al.~\cite{Clayton} using $\Lambda$ produced in interactions of low-energy charged kaon beams with a fixed target, and the measurement of Zech et al.~\cite{Zech}  in 1977 using a neutral hyperon beam. These results are based on data samples containing a maximum of fifty-three thousand events. The relative difference between the lifetimes of $\Lambda$ and $\overline{\Lambda}$ reported in the PDG is $(\tau_{\Lambda}-\tau_{\overline{\Lambda}})/\tau_{\Lambda} = -0.001 \pm 0.009$, resulting from the average of two measurements, one performed in 1967 by Badier et al.~\cite{Badier} and another one in 1996 by Barnes et al.~\cite{Barnes}, using $\Lambda$ and  $\overline{\Lambda}$ produced in low-energy $\mathrm{p} + \mathrm{\overline{p}} \rightarrow \Lambda + \overline{\Lambda}$ reactions. The excellent tracking and particle-identification capabilities of ALICE over a broad momentum range and the large amount of data collected during Run 2 of the LHC are exploited to improve the current precision on the measurement of the $\Lambda$ lifetime and on the relative difference between the lifetimes of $\Lambda$ and $\overline{\Lambda}$. The latter provides a fundamental test of CPT invariance in the strangeness sector. 

This measurement is also a fundamental reference for the studies of the properties of hypernuclear states created in heavy-ion collisions and for future precision studies of other hyperon properties. The analysis presented here is performed in Pb--Pb collisions at a center-of-mass energy \snn $=$ 5.02 TeV using the same data sample employed for the measurements of the (anti)hypertriton lifetime and $\Lambda$ separation energy~\cite{hypertriton_lifetime_Blambda}.  
The latter measurements are fundamental to infer the internal structure of this hypernucleus as well as the properties of hyperon--nucleon interaction in the low-density limit, as described in~\cite{ALICE:2022wpn}.

\section{Experimental apparatus}
\label{sec:Experiment}

ALICE is one of the four large experiments at the LHC and it is dedicated to the study of heavy-ion collisions at ultrarelativistic energies. A detailed description of the ALICE apparatus and its performance can be found in Refs.~\cite{ALICEexperiment} and~\cite{ALICEperformance}. In the following, only the subdetector systems used for the analysis presented in this paper are described.

Trajectories of charged particles are reconstructed in the ALICE central barrel with the Inner Tracking System (ITS)~\cite{ITS} and the Time Projection Chamber (TPC)~\cite{TPC}. These are located within a large solenoidal magnet, providing a highly homogeneous magnetic field of 0.5 T parallel to the beam axis. 
The ITS consists of six cylindrical layers of silicon detectors, concentric and coaxial to the beam pipe, with a total pseudorapidity coverage $|\eta| < 0.9$ with respect to the nominal interaction point. Three different technologies are used for this detector: the two innermost layers consist of silicon pixel detectors (SPD), the two central layers of silicon drift detectors (SDD), and the two outermost layers of double-sided silicon strip detectors (SSD). This detector is used in the determination of primary and secondary vertices, and in the track reconstruction.

The TPC is the largest detector in the ALICE central barrel, with a pseudorapidity coverage $|\eta| < 0.9$. It is used for charged-particle track reconstruction, momentum measurement, and particle identification (PID) via the measurement of the specific energy loss ($\mathrm{d}E/\mathrm{d}x$) of particles in the TPC gas. This detector provides up to 159 spacial points per track for charged-particle reconstruction. The resolution in the measurement of the distance-of-closest approach of primary tracks to the primary collision vertex, projected on the transverse plane, ranges from about 200 $\mu$m at 0.2 GeV$/\textit{c}$ to about 10 $\mu$m at 10 GeV$/\textit{c}$~\cite{ALICEperformance}. The transverse-momentum ($p_{\mathrm{T}}$) resolution ranges from about 1$\%$ at 1 GeV$/\textit{c}$ to about 10$\%$ at 50 GeV$/\textit{c}$ in \PbPb collisions at \snn $=$ 5.02 TeV~\cite{ptSpectraRAA}. The \dEdx resolution depends on the event multiplicity and is about 5--6.5$\%$ for minimum-ionizing particles crossing the full volume of the TPC~\cite{ALICEperformance}. 

The PID is complemented by the Time-Of-Flight (TOF) system~\cite{TOF}. This detector is made of Multi-gap Resistive Plate Chambers (MRPC) and is located at a radial distance of 3.7 m from the nominal interaction point. The TOF detector measures the arrival time of particles relative to the event collision time provided by the TOF detector itself or by the T0 detectors, two arrays of Cherenkov counters located at forward and backward rapidities~\cite{CollisionTime}. The TOF detector is used in this analysis for pile-up rejection, mostly from out-of-bunch collisions, by requiring that at least one of the $\Lambda$ ($\overline{\Lambda}$) charged decay-daughter tracks has an associated hit in the TOF detector. 

Collision events are triggered by two plastic scintillator arrays, V0A and V0C~\cite{VZEROPerformance}, located on both sides of the interaction point, covering the pseudorapidity regions \mbox{2.8 $< \eta <$ 5.1} and $-3.7 <  \eta < -1.7$, respectively. Each array consists of four concentric rings, each ring comprising eight cells with the same azimuthal coverage. The V0A and V0C scintillators are used to determine the collision centrality from the measured signals produced by charged particles~\cite{CentralityDetermination,CentralityDeterminationPublicNote}. The centrality is defined in terms of percentiles of the
total hadronic cross section.       
\section{Data analysis}
\label{sec:DataAnalysis}

\subsection{Event selection}
\label{subsec:EventSelection}

The data used for this analysis were collected in 2018 during the LHC \PbPb run at \snn $=$ 5.02 TeV. A minimum bias (MB) event trigger and two centrality triggers were used. The MB trigger, fully efficient in the centrality interval 0--90$\%$, requires coincident signals in the V0 detectors, synchronous with the bunch crossing time defined by the LHC clock. The two centrality triggers, fully efficient in the centrality classes 0--10$\%$ and 30--50$\%$, are based on the signal amplitude measured by the V0 scintillators, which is proportional to the charged-particle multiplicity of the event. 
The analysis is performed in four centrality classes: 0--10$\%$, 10--30$\%$, 30--50$\%$, and 50--90$\%$. The events in the centrality classes 0--10$\%$ and 30--50$\%$ are selected using both the MB and centrality triggers, while the MB event trigger alone is used for the other centrality classes.
In order to keep the conditions of the detectors as uniform as possible and reject background collisions, the coordinate of the primary vertex along the beam axis is required to be within 10 cm from the nominal interaction point. Events with multiple vertices identified with the SPD are tagged as pile-up and removed from the analysis~\cite{ALICEperformance}. In addition, events with pile-up occurring during the drift time of the TPC are rejected based on the correlation between the number of SDD and SSD clusters and the total number of clusters in the TPC, as described in Ref.~\cite{JPsiFlow}. To further suppress the pile-up contribution, mostly from out-of-bunch collisions, the $\Lambda$ daughter tracks are required to have an associated hit in the TOF detector. This requirement is applied only for the centrality classes 30--50$\%$ and 50--90$\%$. For the most central events, the matching of daughter tracks with a TOF hit 
does not have a significant impact on the fraction of $\Lambda$ ($\overline{\Lambda}$) from events with out-of-bunch pile-up, which is found to be between 0.02$\%$ and 0.07$\%$.  The total number of events selected for each centrality class is reported in Table~\ref{Table:NumberOfEvents}.

\begin{table}[!hbt]
\begin{center}
\centering
\renewcommand{\arraystretch}{1.2}
\caption{Number of events used for the analysis. }
\begin{tabular}{lcclc}
\hline
Centrality                &    &    Number of events ($\times 10^6$)   \\
\hline
       0--10$\%$         &   &     70.97   \\
     10--30$\%$         &   &     18.43   \\
    30--50$\%$         &   &     61.43   \\
     50--90$\%$        &   &     36.86   \\
\hline
\end{tabular}
\label{Table:NumberOfEvents}
\end{center}
\end{table} 


\subsection{Selection of \texorpdfstring{$\Lambda$}{} candidates}
\label{subsec:TrackSelectionPID}

The two-body decay channels $\Lambda \rightarrow \mathrm{p}+\pi^{-}$ and $\overline{\Lambda} \rightarrow \overline{\mathrm{p}}+\pi^{+}$ are used in this measurement. These have a branching ratio BR of $(63.9 \pm 0.5)\%$~\cite{PDG}. The $\Lambda$ ($\overline{\Lambda}$) candidates are reconstructed using the standard ALICE weak decay finder. This algorithm searches for weak decay topologies, called $\mathrm{V}^0$, by reconstructing oppositely-charged particle tracks originating from a displaced vertex as described in Refs.~\cite{LambdaALICE_900GeV,Lambda_K0S_PbPb}. In the case of a decay vertex located inside the ITS volume, at least one hit in any of the ITS layers is used in the reconstruction of the charged tracks originating from the $\mathrm{V}^0$ decay.
The reconstructed tracks, selected in the pseudorapidity region $|\eta| < 0.8$, are required to fulfil a set of quality criteria such as having a number of TPC crossed rows larger than 80, a number of TPC clusters used for the \dEdx calculation larger than 60 to ensure a good \dEdx resolution, a fraction of TPC crossed rows and findable clusters larger than 70$\%$, and a good track fit $\chi^{2}/N^{\rm TPC}_{\rm cls} < 2.5$, where $N^{\rm TPC}_{\rm cls}$ is the number of TPC clusters. 

To reduce the combinatorial background, a set of topological selections are applied, i.e.~the distance of closest approach (DCA) between the $\mathrm{V}^0$ daughter tracks is required to be less than 1~cm, the DCA between the $\mathrm{V}^0$ and the primary collision vertex less than  0.5~cm, the radial distance between primary and secondary vertices larger than 3~cm, and $\mathrm{cos}(\theta_{\rm p}) > 0.995$, where $\theta_{\rm p}$ is the angle between the vector connecting the primary and secondary vertices and the total $\mathrm{V}^0$ momentum ($\vec{p}_{\mathrm{V}^0} = \vec{p}_{\rm p} + \vec{p}_{\pi}$). The selection criteria applied for this measurement are similar to those already used in previous measurements~\cite{Lambda_K0S_PbPb,Hyperons_pp,StrangenessEnhancementNature}.

The particle identification is based on the energy loss per unit of track length measured by the TPC. Protons and pions are identified by requiring that their measured \dEdx is within 3$\sigma_{\mathrm{d}E/\mathrm{d}x}$ from the expected average calculated using the Bethe–Bloch, where $\sigma_{\mathrm{d}E/\mathrm{d}x}$ is the \dEdx resolution. Proton and pion candidates are selected in the transverse-momentum intervals $0.2 < p^{\pi}_{\rm T} < 2 \ \mathrm{GeV}/c$ and $0.2 < p^{\rm p}_{\rm T} < 10 \ \mathrm{GeV}/c$, respectively. 


\subsection{Signal extraction}
\label{subsec:SignalExtraction}

\begin{figure}[hbt]
    \begin{center}
    \includegraphics[width = 0.49\textwidth]{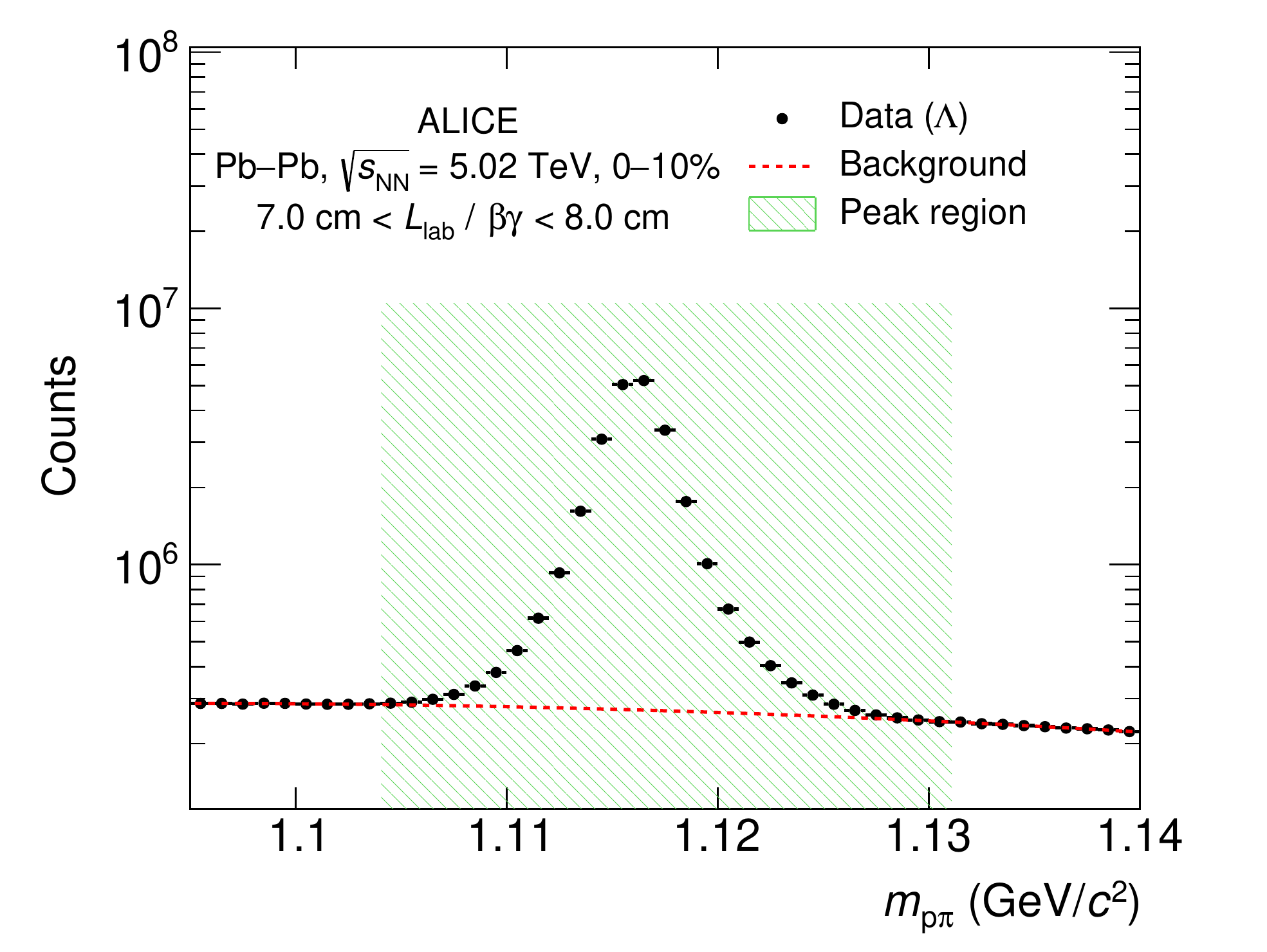}
    \includegraphics[width = 0.49\textwidth]{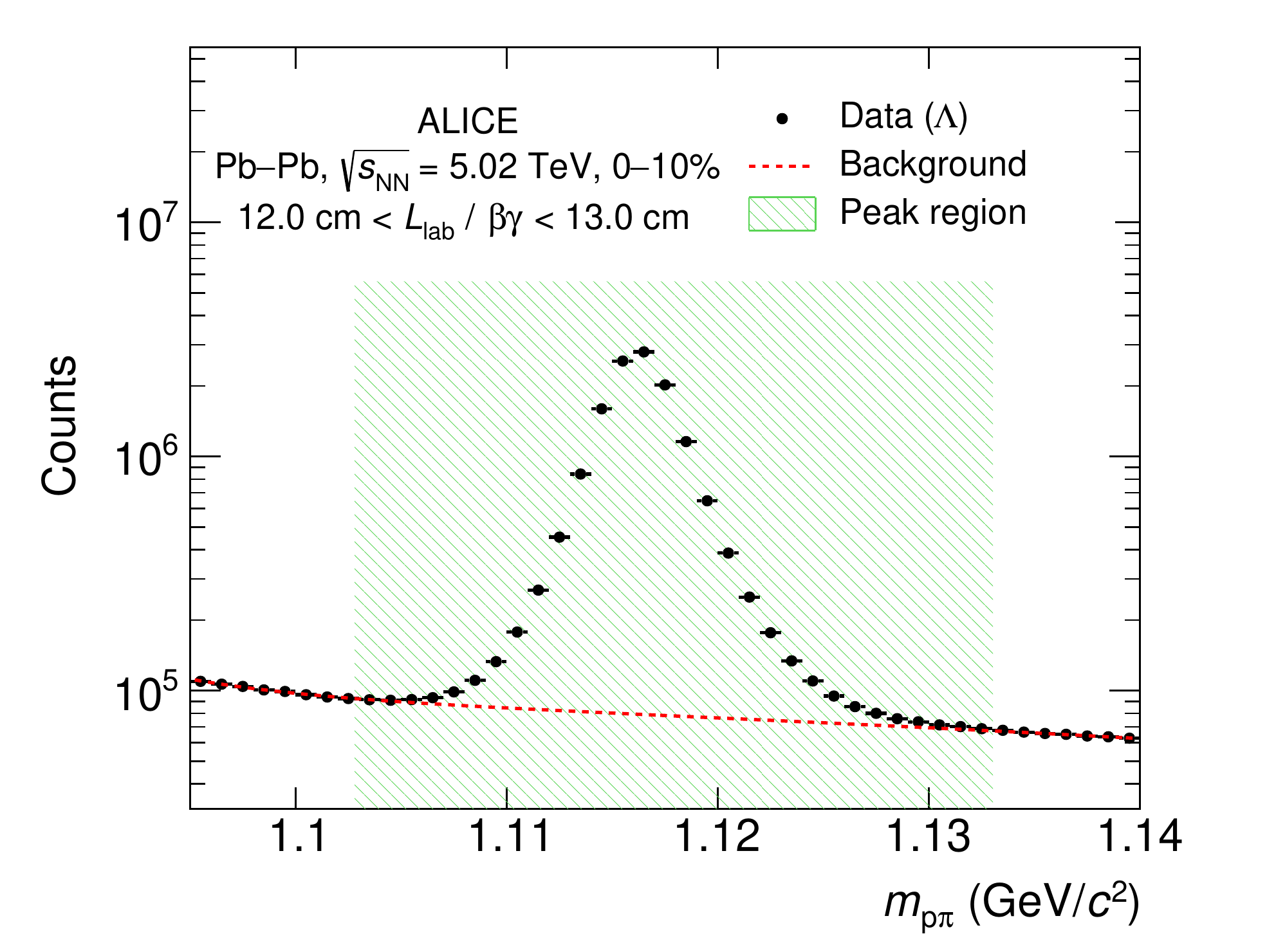}
    \end{center}
\caption{Invariant mass spectra of p$\pi$ pairs measured in central collisions (0--10$\%$) at low (left) and large (right) $L_{\rm proper}$. The green area indicates the peak region. }
\label{fig:signal_extraction}        
\end{figure}

The $\Lambda$ and $\overline{\Lambda}$ lifetimes are extracted from a fit to their proper decay length distributions using the exponential function $\mathrm{exp}\left( -L_{\rm proper}/ \langle L_{\rm proper} \rangle \right)$. The proper decay length is calculated for every $\Lambda$ ($\overline{\Lambda}$) as

\begin{equation}
L_{\rm proper} = L_{\rm lab}/\beta \gamma = M_{\Lambda} \frac{ L_{\rm lab}}{p},
\end{equation}

where $L_{\rm lab}$ is the decay length measured in the laboratory system as the distance between primary and secondary vertices, $M_{\Lambda}$ is the $\Lambda$ mass taken from the PDG ($M_{\Lambda} = 1115.683 \ \mathrm{MeV}/c^2$)~\cite{PDG}, and $p$ is the total momentum of the $\Lambda$ ($\overline{\Lambda}$) measured at the decay point. The number of signal counts in each $L_{\rm proper}$ interval is obtained using the following procedure, which is illustrated in Fig.~\ref{fig:signal_extraction}: 

\begin{enumerate}{}  
\item \textbf{Location of the peak region}: the invariant mass of the decay daughters is calculated and the region around the maximum of the invariant-mass distribution is fitted using a Gaussian function. The peak region is defined as $[M_{0}-8\sigma, M_{0}+10\sigma]$, where $M_{0}$ is the mean and $\sigma$ is the standard deviation of the Gaussian fit. The choice of such a wide signal region is motivated by the fact that the peak has two long tails and is slightly asymmetric, especially for low values of $L_{\rm proper}$. This asymmetry is an effect of residual imperfections in tracking and energy loss corrections, which affect candidates with invariant masses above and below the expected mass differently. 
 
\item \textbf{Fit of the background}: the background on the side bands of the peak is fitted using a continuous function to extrapolate the expected background inside the peak region. Since the background shape changes with $L_{\rm proper}$, a third-order polynomial function is used to fit the background at low $L_{\rm proper}$ while the sum of a power-law and a linear function is used at high $L_{\rm proper}$ values. These functions have the minimum number of parameters that guarantees a data-to-fit ratio consistent with unity within statistical uncertainties in the sidebands. 

\item \textbf{Signal extraction}: the signal is extracted in each $L_{\rm proper}$ interval by subtracting the estimated background from the invariant mass distribution and counting the entries inside the peak region. 

\end{enumerate}

The total number of $\Lambda$ and $\overline{\Lambda}$ raw counts within the peak region are reported in Table~\ref{Table:RawCounts} for each centrality interval.

\begin{table}[!hbt]
\begin{center}
\centering
\renewcommand{\arraystretch}{1.2}
\caption{Raw counts of $\Lambda$ and $\overline{\Lambda}$ for different centralities. }
\begin{tabular}{lcclc}
\hline
Centrality                &   $\Lambda$ ($\times 10^6$)   &    $\overline{\Lambda}$ ($\times 10^6$)   \\
\hline
       0--10$\%$         &  312.6  &     296.4    \\
     10--30$\%$         & 49.7  &     47.2   \\
    30--50$\%$         &  41.2 &     35.9   \\
     50--90$\%$        &  4.2 &    3.6     \\
\hline
\end{tabular}
\label{Table:RawCounts}
\end{center}
\end{table}


\subsection{Efficiency and secondary \texorpdfstring{$\Lambda$}{} corrections}
\label{subsec:EfficiencyFeeddown}

The raw $L_{\rm proper}$ spectrum of $\Lambda$ ($\overline{\Lambda}$) is corrected for the reconstruction efficiency, the feed-down contribution from higher-mass baryons  and secondary $\Lambda$ ($\overline{\Lambda}$) originating from interactions of particles with the detector material as

\begin{equation}
\left[ \frac{\mathrm{d}N_{\Lambda}}{\mathrm{d}L_{\rm proper}} \right]_{\rm corr} = \frac{1}{\epsilon (L_{\rm proper})} \times  f_{\rm prim} (L_{\rm proper})  \times \left[ \frac{\mathrm{d}N_{\Lambda}}{\mathrm{d}L_{\rm proper}} \right]_{\rm raw}, 
\end{equation}

where $\epsilon (L_{\rm proper})$ is the efficiency of primary $\Lambda$ ($\overline{\Lambda}$) and $f_{\rm prim}  (L_{\rm proper})$ is the fraction of primary $\Lambda$ ($\overline{\Lambda}$). The dominant feed-down contributions are given by the weak decays of $\Xi^{\pm}$, $\Xi^{0}$, and $\Omega^{\pm}$:~\cite{PDG}

\begin{list}{$\bullet$}{}  
\item $\Xi^{0} (\overline{\Xi}^{0}) \rightarrow \Lambda (\overline{\Lambda}) + \pi^{0}$                     \ \ \ BR $=$ $(99.524 \pm 0.012)\%$, 
\item $\Xi^{\pm} \rightarrow \Lambda + \pi^{\pm}$             \ \ \ BR $=$ $(99.887 \pm 0.035)\%$, 
\item $\Omega^{\pm} \rightarrow \Lambda + K^{\pm}$      \ \ \ BR $=$ $(67.8 \pm 0.7)\%$,  
\item $\Omega^{\pm} \rightarrow  \Xi^{0} + \pi^{\pm} \ \mathrm{and} \ \Omega^{\pm} \rightarrow \Xi^{\pm} + \pi^0$      \ \ \ BR $=$ $(32.2 \pm 0.8)\%$.
\end{list}

These corrections are calculated using Monte Carlo (MC) simulations. Collision events between lead ions are simulated using the HIJING event generator~\cite{HIJING} and the passage of particles through the experimental apparatus is simulated using GEANT3~\cite{GEANT3} as transport code. Considering that the $p_{\rm T}$ distributions of particles and their relative abundances in MC simulations are different from data, centrality and $p_{\rm T}$-dependent corrections are applied in MC simulations using weights. 
These are defined, for different centrality classes, as the ratio of the $p_{\rm T}$ spectrum measured by ALICE and the $p_{\rm T}$  spectrum generated by HIJING.

The ALICE measurements of the \pt spectra of $\Lambda$, $\Xi^{\pm}$, and $\Omega^{\pm}$~\cite{Lambda_K0S_PbPb,XiOmegaALICE_276TeV} in Pb--Pb collisions at \snn $=$ 2.76~TeV, scaled by the ratio of the proton \pt spectra measured at \snn $=$ 5.02~TeV~\cite{protons_502TeV} and \snn $=$ 2.76~TeV~\cite{protons_276TeV}, are used to calculate the weights for different centralities. Based on isospin symmetry, the \pt spectra of $\Xi^{0}$ are assumed to be identical to those of $\Xi^{\pm}$. Centrality dependent factors, given by the ratios $(\Xi/\Lambda)_{\rm data}/ (\Xi/\Lambda)_{\rm MC}$ and $(\Omega/\Lambda)_{\rm data}/ (\Omega/\Lambda)_{\rm MC}$, are also included to reproduce the centrality dependence of the particle ratios observed in data. The efficiency is calculated as the ratio between reconstructed and generated primary $\Lambda$ in the simulation

\begin{equation}
\epsilon (L_{\rm proper}) =  \frac{\left[ \frac{\mathrm{d}N_{\Lambda}}{\mathrm{d}L_{\rm proper}} \right]_{\rm rec}}{ \left[ \frac{\mathrm{d}N_{\Lambda}}{\mathrm{d}L_{\rm proper}} \right]_{\rm gen} }.
\end{equation}

\begin{figure}[hbt]
    \begin{center}
    \includegraphics[width = 0.49\textwidth]{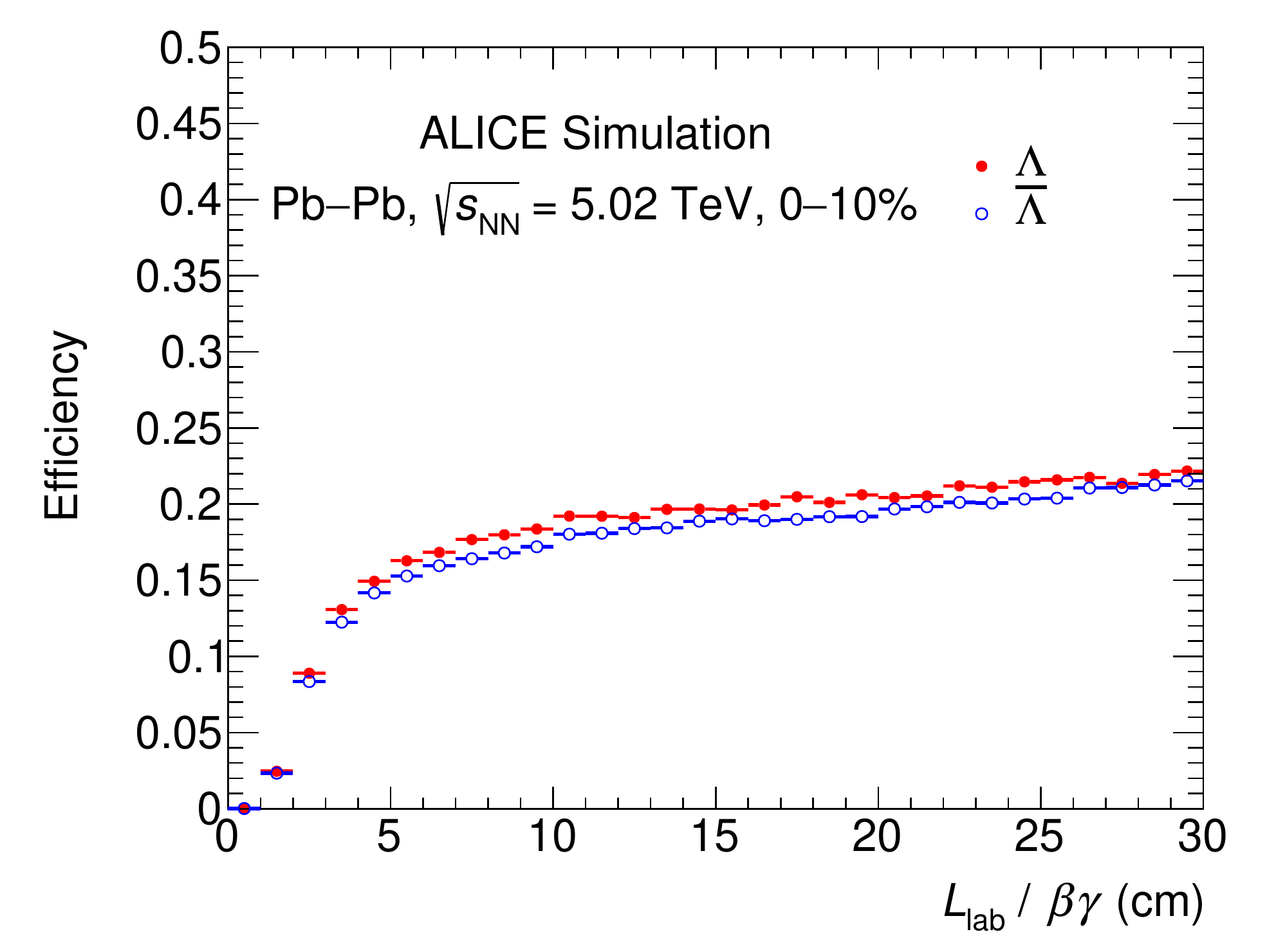}
    \includegraphics[width = 0.49\textwidth]{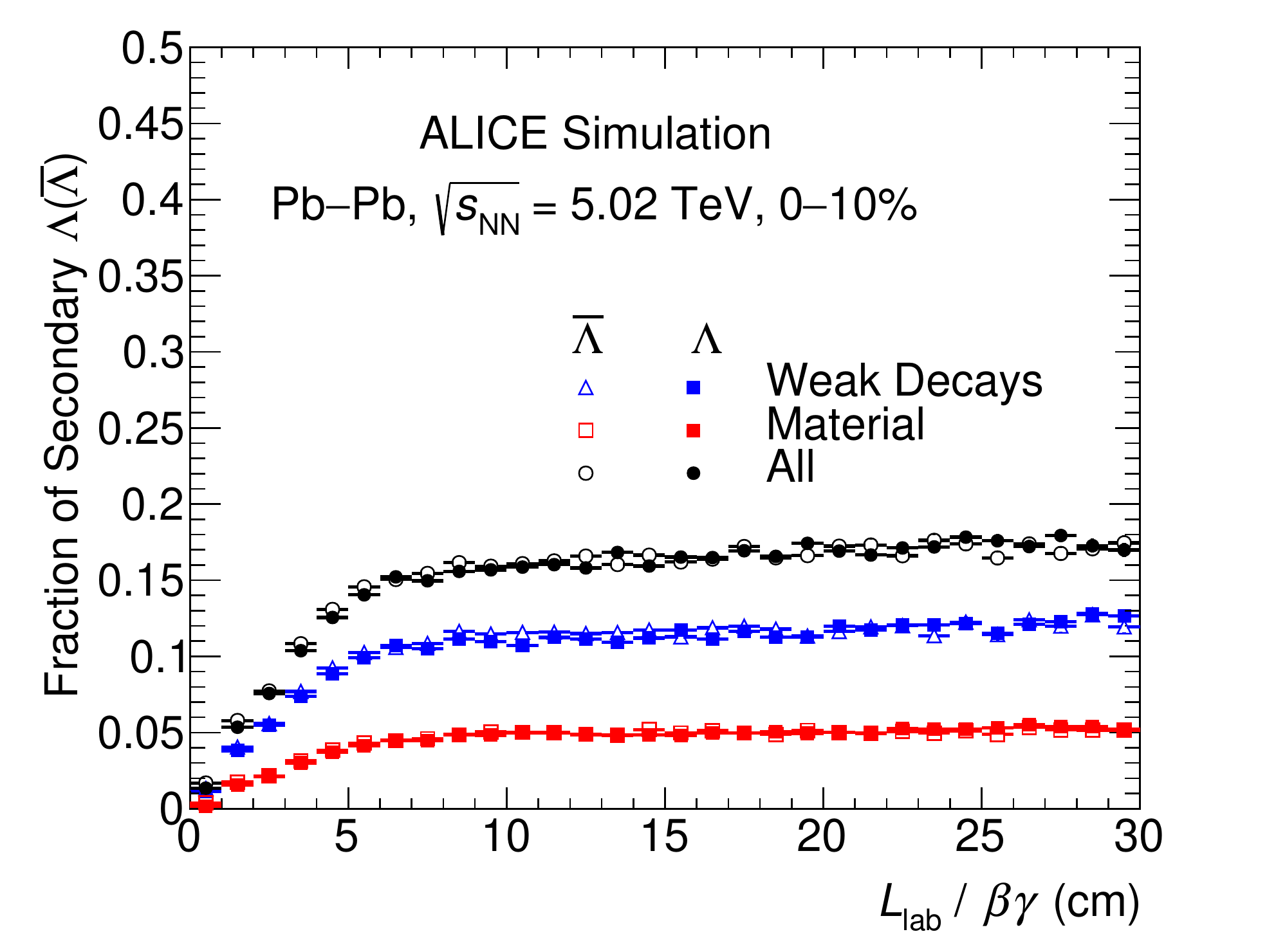}
    \end{center}
\caption{Reconstruction efficiency of primary $\Lambda$ ($\overline{\Lambda}$) (left) and fraction of secondary $\Lambda$ ($\overline{\Lambda}$) (right) in central Pb--Pb collisions (0--10$\%$). }
\label{fig:corrections}        
\end{figure}

The efficiencies of $\Lambda$ and $\overline{\Lambda}$ for central Pb--Pb collisions (0--10$\%$) as a function of $L_{\rm proper}$ are shown in Fig.~\ref{fig:corrections} (left). The correction for secondary $\Lambda$ ($\overline{\Lambda}$) from material and weak decays is applied by scaling the raw $L_{\rm proper}$ spectrum by the fraction of primary $\Lambda$ ($\overline{\Lambda}$) given by

\begin{equation}
f_{\rm prim} (L_{\rm proper}) = 1 - \frac{ \mathrm{d}N_{\Lambda_{\rm sec}}/ \mathrm{d}L_{\rm proper}}{\mathrm{d}N_{\Lambda_{\rm all}}/ \mathrm{d}L_{\rm proper}}  .
\end{equation}

The fraction of secondary $\Lambda$ and $\overline{\Lambda}$ for central Pb--Pb (0--10$\%$) is shown as a function of $L_{\rm proper}$ in the right panel of Fig.~\ref{fig:corrections}. The individual contributions from material and weak decays are shown in addition to the total fraction for illustration. The observed trend of the fraction of secondary $\Lambda$ ($\overline{\Lambda}$) from weak decays with $L_{\rm proper}$ is due to an interplay between the efficiency and decay time of the $\Lambda$ ($\overline{\Lambda}$) mother particle. For secondary $\Lambda$ ($\overline{\Lambda}$) from material, it is due to a combined effect of efficiency and the radial distance at which the secondary $\Lambda$ ($\overline{\Lambda}$) is produced.

\section{Systematic uncertainties}
\label{sec:SystematicUncertainties}

The dominant sources of systematic uncertainties on the $\Lambda$ and $\overline{\Lambda}$ lifetime measurements are related to the track and $\mathrm{V}^0$ selections, signal extraction, efficiency and feed-down corrections. These are summarized in Table~\ref{Table:syst_uncertainty_summary}. The methods used to estimate the systematic uncertainties from these sources are illustrated in the following. In addition, effects of the material budget uncertainty, the uncertainty on the hadronic interaction of $\Lambda$, $\overline{\Lambda}$, and their decay daughters, and potential effects of residual pile-up \textemdash which are found to be negligible \textemdash are also discussed.

\begin{table}[!hbt]
\begin{center}
\centering
\renewcommand{\arraystretch}{1.1}
\caption{Summary of the systematic uncertainties on the $\Lambda$ and $\overline{\Lambda}$ lifetime measurements. All values are in ps.}
\begin{tabular}{lclclcl}
\hline
Source      &         $\Lambda$      &       \ \ $\overline{\Lambda}$    &    $\Lambda$ + $\overline{\Lambda}$  \\
\hline
\hline
Track and $\mathrm{V}^0$ selections     &   0.55    &    0.69  &     0.65   \\
Signal extraction     &   0.03      &   0.03      &   0.02     \\
Efficiency and feed-down corrections   &    0.30   &    0.33 &     0.30  \\
\hline
Total     &   0.63    &    0.77  &     0.72   \\
\hline
\end{tabular}
\label{Table:syst_uncertainty_summary}
\end{center}
\end{table}

\subsection{Systematic uncertainty from track and \texorpdfstring{$\mathrm{V}^0$}{} selection}
\label{subsec:TrackV0Uncertainty}

The systematic uncertainty due to the track and $\mathrm{V}^0$ selection criteria is estimated by repeating the full analysis chain using one hundred different analysis parameters, where the single-track, topological, and particle-identification selection criteria are varied, 
such that they produce a maximum variation of $\pm 10\%$ in the raw signal yield, similarly to the approach used in Refs. \cite{Lambda_K0S_PbPb,Hyperons_pp,StrangenessEnhancementNature}.
The systematic uncertainty from the track and $\mathrm{V}^0$ selection is calculated by fitting the distribution of lifetime values obtained from different selection criteria using a Gaussian function and taking the $\sigma$ of the Gaussian fit as an uncertainty. The obtained uncertainties are 0.55 ps for $\tau_{\Lambda}$, 0.69 ps for $\tau_{\overline{\Lambda}}$, and 0.65 ps for $\tau_{\Lambda + \overline{\Lambda}}$. This contribution is the dominant source of systematic uncertainty.

\subsection{Signal extraction uncertainty}
\label{subsec:SignalExtractionUncertainty}

The systematic uncertainty from the signal extraction includes two contributions: the choice of the background fit range and the integration range used for the raw yield extraction. For both contributions, one hundred different intervals are randomly generated, with a uniform probability distribution between two extremes, and the signal extraction procedure is repeated for each of these intervals. The limits used are specified in Table~\ref{Table:extremes}.

\begin{table}[thb]
\begin{center}
\centering
\renewcommand{\arraystretch}{1.1}
\caption{Invariant-mass intervals used for the signal extraction systematic uncertainty.}
\begin{tabular}{lclclcl}
\hline
       &         Left extreme      &       Right extreme      \\
\hline
\hline
Background fit range     &   $[M_{0}-14\sigma, M_{0}-11\sigma]$      &   $[M_{0}+20\sigma, M_{0}+32\sigma]$     \\
Signal integration range     &   $[M_{0}-13\sigma, M_{0}-7\sigma]$      &   $[M_{0}+5\sigma, M_{0}+11\sigma]$   \\
\hline
\end{tabular}
\label{Table:extremes}
\end{center}
\end{table}

The standard deviation of the distribution of raw yields in each $L_{\rm proper}$ interval is taken as the systematic uncertainty. These two contributions are independent and added in quadrature. The stability of the results was tested against different choices of the fit functions used to model the background. In particular, linear and second-order polynomials were also tried at low $L_{\rm proper}$, while an exponential (power-law) for the left and second-order polynomial (exponential) for the right side of the background were also used at high $L_{\rm proper}$. The use of alternative fit functions to model the background resulted in negligible changes in the extracted yield. For this reason, this contribution to the signal extraction uncertainty is considered negligible. The uncertainty on the lifetime is calculated by replacing the statistical uncertainties on the corrected $L_{\rm proper}$ spectrum with the signal extraction uncertainties, which are bin-by-bin uncorrelated, and taking the uncertainty from the exponential fit as the systematic uncertainty. This contribution is found to be $(\Delta\tau)^{\rm syst}_{\rm signal} = 0.03$ ps for both $\tau_{\Lambda}$ and $\tau_{\overline{\Lambda}}$, and 0.02 ps for $\tau_{\Lambda + \overline{\Lambda}}$.

\subsection{Systematic uncertainty from efficiency and feed-down corrections}
\label{subsec:EfficiencyUncertainty}

The systematic uncertainty related to the efficiency and feed-down corrections is estimated by varying: (i) the $p_{\rm T}$-dependent weights used to adjust the input $p_{\rm T}$ distributions of $\Lambda$, $\Xi$, and $\Omega$ in the simulations, (ii) the $\Lambda$, $\Xi$, and $\Omega$ lifetimes by the PDG uncertainties \cite{PDG} and (iii) the $\Lambda/\Xi$ and $\Lambda/\Omega$ ratios by the measured uncertainties. The $\Omega$ is found to give a negligible contribution to the systematic uncertainties.
To vary the lifetimes implemented in the simulations, which are taken from the PDG \cite{PDG}, $L_{\rm proper}$-dependent weights are used. These are obtained as the ratio between the $L_{\rm proper}$ spectrum with modified lifetime and the default spectrum. To estimate the total contribution of these sources, a set of five hundred different efficiency and feed-down corrections is generated. Each of them is obtained using a different set of weights where the  $p_{\rm T}$ spectra of $\Lambda$, $\Xi$, and $\Omega$, their lifetimes and particle ratios are varied by a fraction of their uncertainty. Such a fraction is extracted randomly from a Gaussian distribution centered at zero and with a width equal to 1. When modifying the $p_{\rm T}$ spectra measured in data to recalculate the weights, the correlated and uncorrelated uncertainties with $p_{\rm T}$ are treated differently: 

\begin{enumerate}{}  
\item  \textbf{$p_{\rm T}$-correlated uncertainties}: all data points of the $p_{\rm T}$ spectrum are shifted coherently upward and downward by a fraction of their systematic uncertainty in each $p_{\rm T}$ interval.
\item  \textbf{$p_{\rm T}$-uncorrelated uncertainties}: the data points are moved independently by a fraction of their uncorrelated uncertainty in each $p_{\rm T}$ interval.
\end{enumerate}

These five hundred different efficiencies and fractions of secondary $\Lambda$ ($\overline{\Lambda}$) are then used to correct the raw-$L_{\rm proper}$ spectrum measured in data. The lifetime is extracted for each corrected spectrum and the standard deviation of the distribution of lifetimes is taken as an estimate of the systematic uncertainty from efficiency and feed-down corrections. The obtained uncertainties are 0.30 ps for $\tau_{\Lambda}$, 0.33 ps for $\tau_{\overline{\Lambda}}$, and 0.30 ps for $\tau_{\Lambda + \overline{\Lambda}}$.


\subsection{Inelastic interaction with the detector materials}
\label{subsec:Absorption}

The default efficiency is based on GEANT3 transport package. The effect of (anti)matter absorption was studied by comparing the default efficiency with that obtained using a MC production based on GEANT4 \cite{GEANT4}, which contains slightly different parametrizations of the inelastic cross sections of (anti)matter particles. The GEANT3 and GEANT4-based efficiencies are consistent within uncertainties. The $\Lambda$ and $\overline{\Lambda}$ lifetimes are found to be consistent within uncertainties, and therefore, no systematic uncertainty is assigned due to this effect.

\subsection{Material budget}
\label{subsec:MaterialBudget}

The ALICE detector material is known with a precision of 4.5$\%$ \cite{ALICEperformance}. The effect of the limited knowledge of the material budget, which could affect the efficiency and the fraction of secondary $\Lambda$ ($\overline{\Lambda}$) from material and its dependence on $L_{\rm proper}$, is studied by comparing the efficiency and corrections for secondary $\Lambda$ ($\overline{\Lambda}$) using MC productions with increased and decreased material density by 4.5$\%$. The difference in the mean lifetime is found to not be statistically significant and therefore this contribution is neglected. 

\subsection{Pile-up effects }
\label{subsec:PileupEffects}

Simultaneous collisions with displaced vertices (pile-up) could, in principle, create a bias in the measurement of the $\Lambda$ ($\overline{\Lambda}$) decay length due to the wrong $\mathrm{V}^0$--vertex association. The tight selection on the $\mathrm{cos}(\theta_{\rm p})$ allows the matching between a reconstructed $\mathrm{V}^0$ and the wrong vertex only for close vertices. This happens with very low probability and is found to give a negligible bias in the decay length measurement. To further cross-check potential pile-up effects, the analysis is repeated removing all pile-up rejections. The $\Lambda$ ($\overline{\Lambda}$) lifetimes, in this case, are found to be consistent with the value using default pile-up rejection within the statistical uncertainties. It is concluded that the pile-up effects combined with rather strong topological selections used in this analysis give a negligible effect on the $\Lambda$ ($\overline{\Lambda}$) lifetime.

\section{Results}
\label{sec:Results}

\begin{figure}[hbt]
    \begin{center}
    \includegraphics[width = 0.99\textwidth]{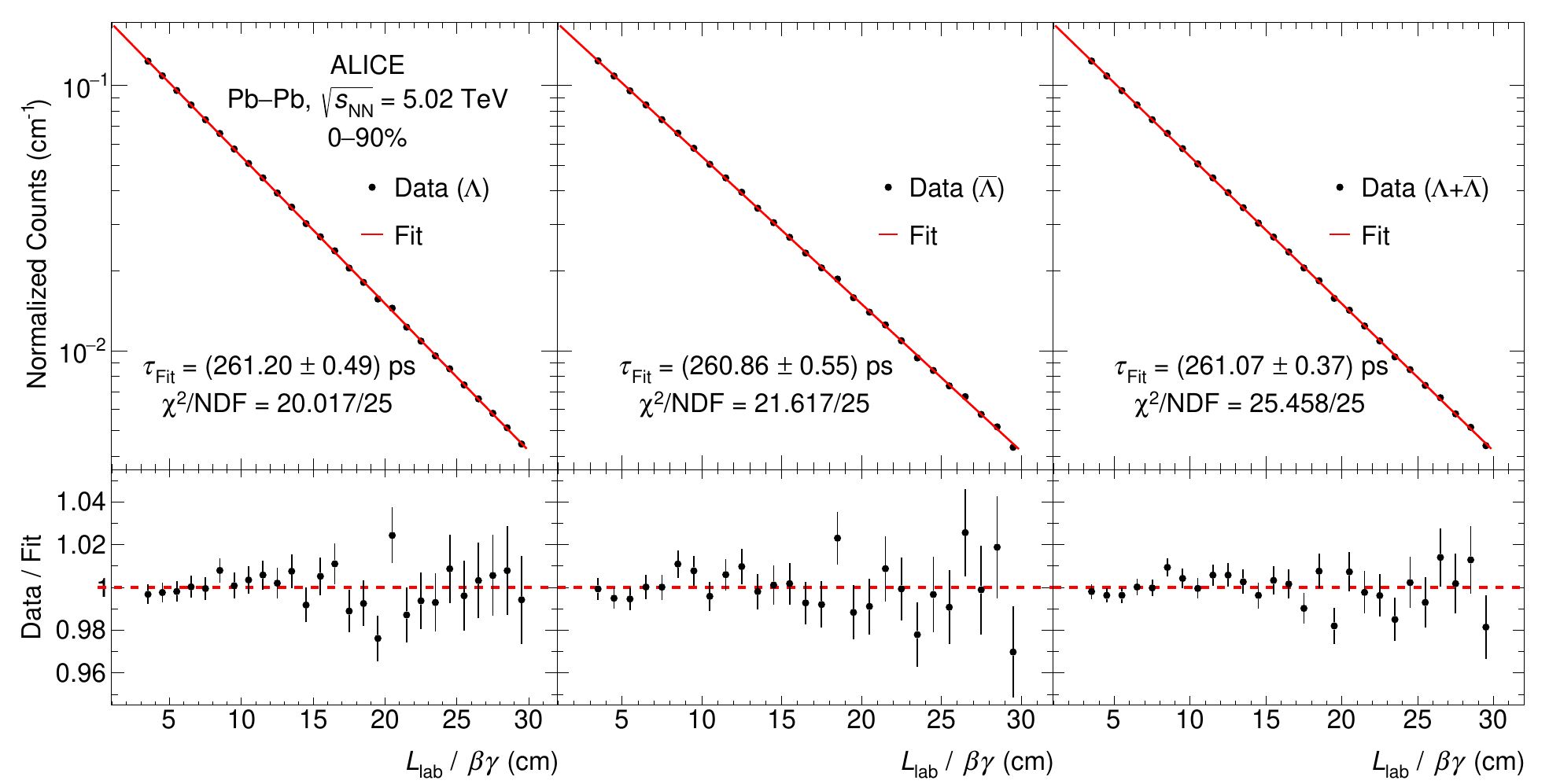}
    \end{center}
\caption{$L_{\rm proper}$ spectra of $\Lambda$ (left), $\overline{\Lambda}$ (middle) and their sum (right), and exponential fits for the lifetime extractions. Only statistical uncertainties are shown for each data point and for the mean lifetime extracted from the exponential fit. }
\label{fig:lifetime_measurement}        
\end{figure}

The $L_{\rm proper}$ spectra of $\Lambda$ and $\overline{\Lambda}$ measured in each centrality interval are corrected for the corresponding efficiency, feed-down from higher mass baryons, and the fraction of secondary $\Lambda$ ($\overline{\Lambda}$) from the material. 
The use of centrality triggers in the data used in this analysis leads to a non-uniform centrality distribution. In order to restore the correct relative contribution from different centralities, the $L_{\rm proper}$ spectra measured in the centrality intervals 10--30$\%$, 30--50$\%$, and 50--90$\%$ are scaled by a factor

\begin{equation}
k_i = \frac{N_{\rm events}^{0-10\%}/w_{0-10\%}}{N_{\rm events}^{i}/w_{i}} w_i,
\label{Eq:spectra_combination}        
\end{equation}

where  $N_{\rm events}^{i}$ and $w_i$ are the number of events and the centrality bin width of the $i$-th centrality interval and $N_{\rm events}^{0-10\%}$ and $w_{0-10\%} = 10$ are those related to the reference centrality interval 0--10$\%$. The obtained spectrum is normalized to unity and fitted with an exponential function in the $L_{\rm proper}$ interval [3,30] cm to extract the mean lifetime. The fit results for $\Lambda$, $\overline{\Lambda}$ and their sum are shown in Fig.~\ref{fig:lifetime_measurement}, which also contains the data-to-fit ratios in the bottom panels. The data-to-fit ratio in each interval is obtained as the ratio between the interval content and the weighted average of the fit function within the interval, with weight given by the exponential function. In this figure, only statistical uncertainties in each $L_{\rm proper}$ interval and on the mean lifetime are shown. The systematic uncertainties are calculated using the procedure described in Sec.~\ref{sec:SystematicUncertainties} and are reported only for the final result of the lifetime in the 0--90$\%$ centrality class. The fit is stable when changing fit range ($L^{\rm min}_{\rm proper}$ $=$ 1,2,3,4,5,..., 10 cm) and binning (width $=$ 0.5,1,2 cm)  leading to results that are consistent within the statistical uncertainties. 

The measured lifetimes of $\Lambda$ and $\overline{\Lambda}$ with statistical and systematic uncertainties, are

\begin{align*}
\tau_{\Lambda} = [261.20 \pm 0.49 (\rm stat.) \pm 0.63 (\rm syst.)]\ \rm ps, \\
\tau_{\overline{\Lambda}} = [260.86 \pm 0.55 (\rm stat.) \pm 0.77 (\rm syst.)]\ \rm ps, \\
\tau_{\Lambda+\overline{\Lambda}} = [261.07 \pm 0.37 (\rm stat.) \pm 0.72 (\rm syst.)]\ \rm ps.
\end{align*}

\begin{figure}[hbt]
    \begin{center}
    \includegraphics[width = 0.6\textwidth]{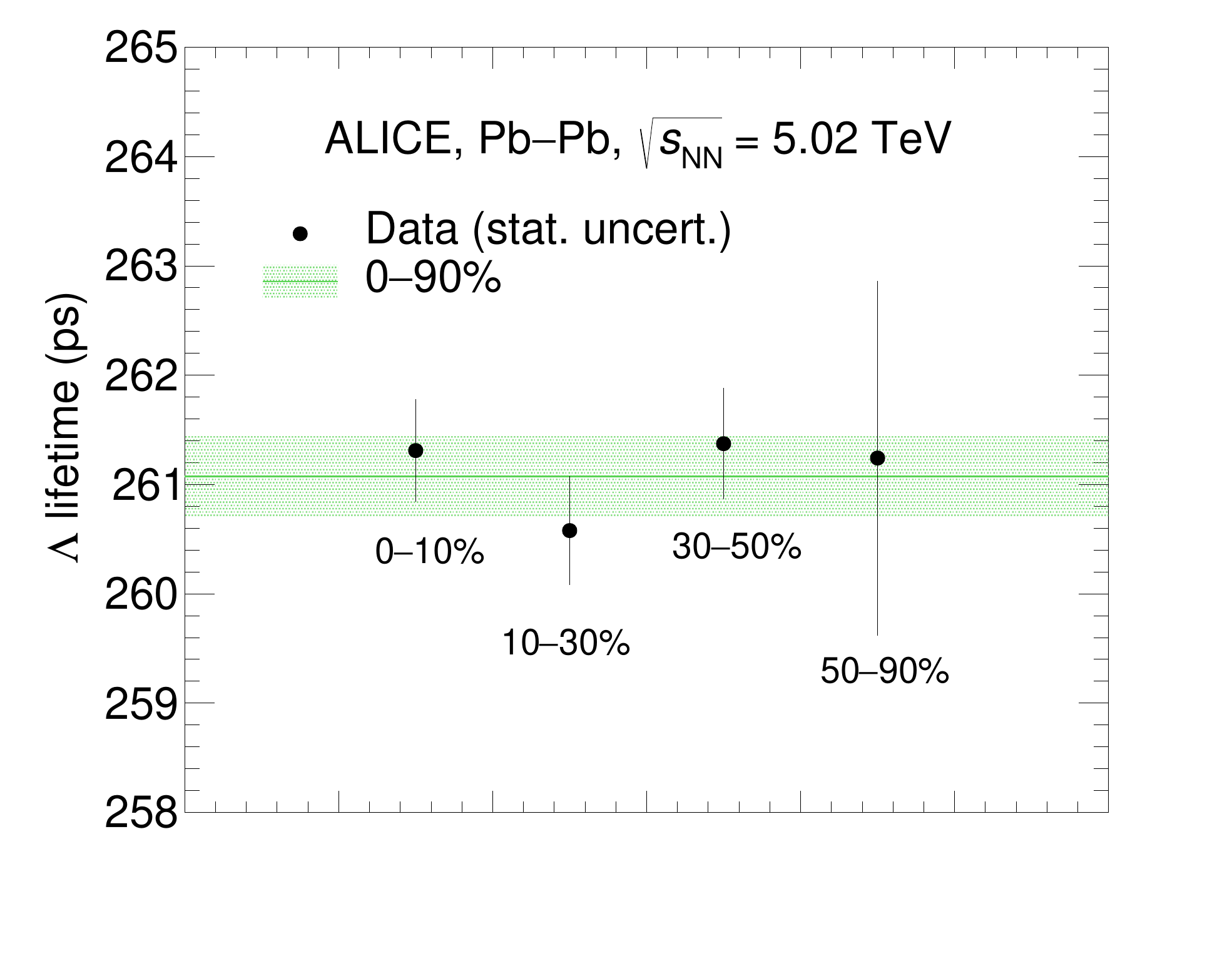}
    \end{center}
\caption{$\Lambda$ lifetime measured in different centrality intervals. Only statistical uncertainties are shown. }
\label{fig:comparison_centralities}        
\end{figure}

The lifetimes extracted in different centrality intervals are consistent within their statistical uncertainties, as shown in Fig.~\ref{fig:comparison_centralities}. As a cross-check, the lifetime is also calculated as the weighted average of the results in different centrality intervals, with weights given by the inverse of the statistical uncertainties squared. The result is fully consistent with that extracted from the $L_{\rm proper}$ distribution obtained using Eq.~\ref{Eq:spectra_combination}.

\begin{figure}[hbt]
    \begin{center}
    \includegraphics[width = 0.6\textwidth]{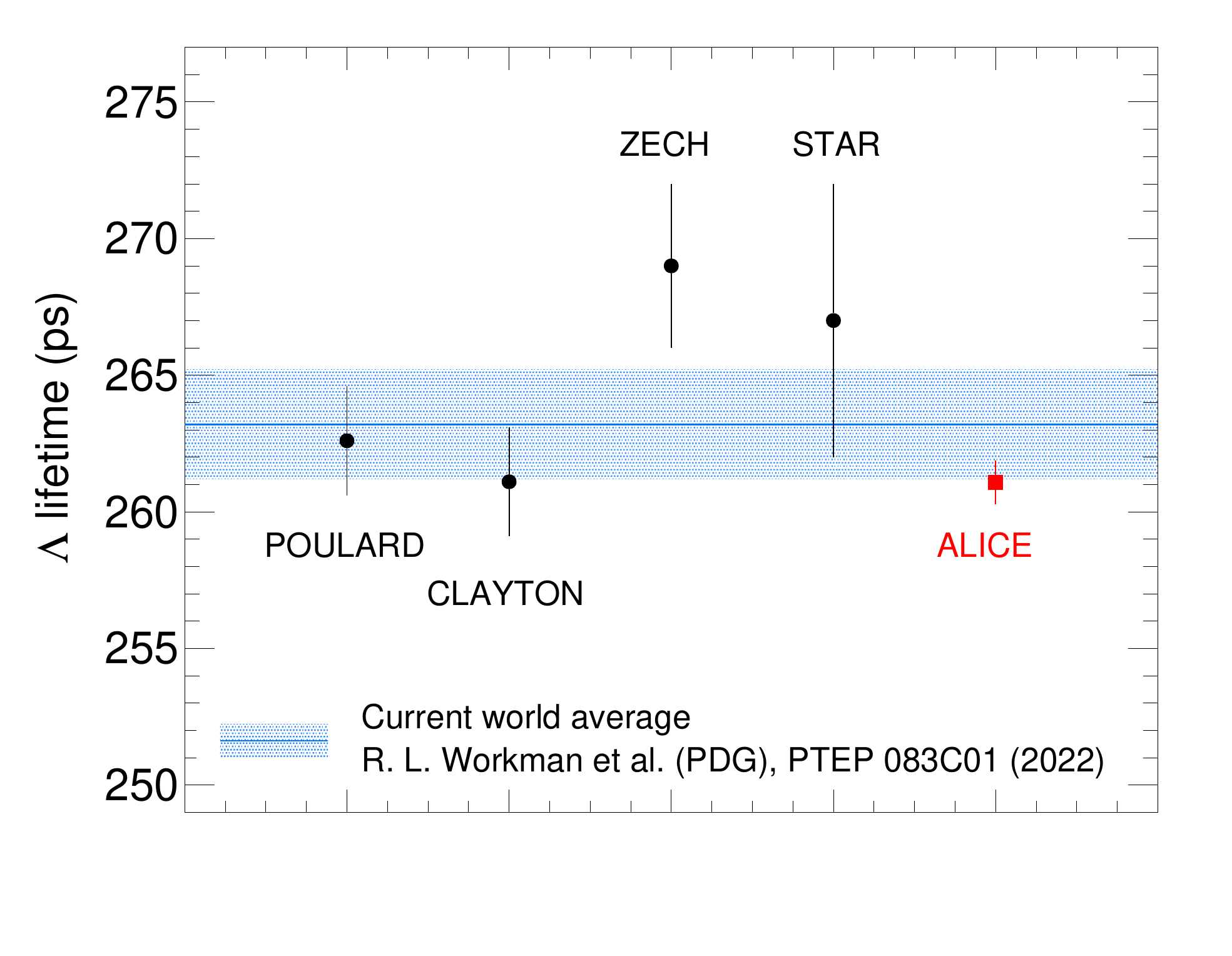}
    \end{center}
\caption{ALICE measurement of the $\Lambda$ lifetime in comparison with previous measurements \cite{Poulard,Clayton,Zech,STARLambdaLifetime} and the current world average taken from the PDG \cite{PDG}. Statistical and systematic uncertainties are added in quadrature. }
\label{fig:comparison_pdg}        
\end{figure}

The present measurement is compared with previous results in Fig.~\ref{fig:comparison_pdg}. The STAR measurement is taken from Ref. \cite{STARLambdaLifetime}. For this comparison, statistical and systematic uncertainties are added in quadrature.

Assuming CPT invariance, the lifetimes of the $\Lambda$ and $\overline{\Lambda}$ are expected to be consistent within uncertainties. To test CPT invariance, the relative difference $( \tau_{\Lambda} - \tau_{\overline{\Lambda}}) /  \tau_{\Lambda}$ is measured. Statistical uncertainties on $\tau_{\Lambda}$ and $\tau_{\overline{\Lambda}}$ and the systematic uncertainties originating from the signal extraction are uncorrelated and propagated independently. On the other hand, the systematic uncertainties on the efficiency and feed-down corrections as well as those on the track and $\mathrm{V}^0$ selections of $\Lambda$ and $\overline{\Lambda}$ are partially correlated. The systematic uncertainty on the relative difference $( \tau_{\Lambda} - \tau_{\overline{\Lambda}}) /  \tau_{\Lambda}$ from the former contribution is considered as half of the interval with the following extremes

\begin{equation}
\left[\frac{\tau_{\overline{\Lambda}}}{ \tau_{\Lambda}} \right]_{\rm upper} =  \frac{\tau_{\overline{\Lambda}} + \Delta \tau_{\overline{\Lambda} } (\rm corrections)}{ \tau_{\Lambda}+\Delta \tau_{\Lambda} (\rm corrections)},  \
\end{equation}

\begin{equation}
\left[\frac{\tau_{\overline{\Lambda}}}{ \tau_{\Lambda}} \right]_{\rm lower} =  \frac{\tau_{\overline{\Lambda}} - \Delta \tau_{\overline{\Lambda}} (\rm corrections) }{ \tau_{\Lambda}-\Delta \tau_{\Lambda} (\rm corrections)}, 
\end{equation}

and is found to be $1.1 \times10^{-4}$. To take into account the correlation between the uncertainties from the track and $\mathrm{V}^0$ selections, the relative difference $( \tau_{\Lambda} - \tau_{\overline{\Lambda}}) /  \tau_{\Lambda}$ is calculated for each of the different analysis settings used. The uncertainty is given by the standard deviation of the distribution of $( \tau_{\Lambda} - \tau_{\overline{\Lambda}}) /  \tau_{\Lambda}$ values and is found to be 0.0021, which is the largest contribution to the total systematic uncertainty.    
The measured value of the relative difference between the $\Lambda$ and $\overline{\Lambda}$ lifetimes is

\begin{equation}
(\tau_{\Lambda} - \tau_{\overline{\Lambda}}) /  \tau_{\Lambda} = 0.0013 \pm 0.0028 (\rm stat.) \pm 0.0021 (\rm syst.),
\end{equation}

while that reported in the PDG is $( \tau_{\Lambda} - \tau_{\overline{\Lambda}}) /  \tau_{\Lambda} = -0.001 \pm 0.009$ \cite{PDG}. The present measurement is consistent with zero with an overall improvement of the absolute precision with respect to the PDG by approximately a factor of three.
       
\section{Summary}
\label{sec:Summary}

Unprecedentedly precise measurements of the $\Lambda$ ($\overline{\Lambda}$) lifetime and of the relative difference between the lifetimes of $\Lambda$ and $\overline{\Lambda}$ are presented. The latter represents an important test of the CPT symmetry in the strangeness sector. The confidence range of the $\Lambda$ ($\overline{\Lambda}$) lifetime is reduced by approximately a factor of three with respect to the PDG average. 

\newenvironment{acknowledgement}{\relax}{\relax}
\begin{acknowledgement}
\section*{Acknowledgements}

The ALICE Collaboration would like to thank all its engineers and technicians for their invaluable contributions to the construction of the experiment and the CERN accelerator teams for the outstanding performance of the LHC complex.
The ALICE Collaboration gratefully acknowledges the resources and support provided by all Grid centres and the Worldwide LHC Computing Grid (WLCG) collaboration.
The ALICE Collaboration acknowledges the following funding agencies for their support in building and running the ALICE detector:
A. I. Alikhanyan National Science Laboratory (Yerevan Physics Institute) Foundation (ANSL), State Committee of Science and World Federation of Scientists (WFS), Armenia;
Austrian Academy of Sciences, Austrian Science Fund (FWF): [M 2467-N36] and Nationalstiftung f\"{u}r Forschung, Technologie und Entwicklung, Austria;
Ministry of Communications and High Technologies, National Nuclear Research Center, Azerbaijan;
Conselho Nacional de Desenvolvimento Cient\'{\i}fico e Tecnol\'{o}gico (CNPq), Financiadora de Estudos e Projetos (Finep), Funda\c{c}\~{a}o de Amparo \`{a} Pesquisa do Estado de S\~{a}o Paulo (FAPESP) and Universidade Federal do Rio Grande do Sul (UFRGS), Brazil;
Bulgarian Ministry of Education and Science, within the National Roadmap for Research Infrastructures 2020-2027 (object CERN), Bulgaria;
Ministry of Education of China (MOEC) , Ministry of Science \& Technology of China (MSTC) and National Natural Science Foundation of China (NSFC), China;
Ministry of Science and Education and Croatian Science Foundation, Croatia;
Centro de Aplicaciones Tecnol\'{o}gicas y Desarrollo Nuclear (CEADEN), Cubaenerg\'{\i}a, Cuba;
Ministry of Education, Youth and Sports of the Czech Republic, Czech Republic;
The Danish Council for Independent Research | Natural Sciences, the VILLUM FONDEN and Danish National Research Foundation (DNRF), Denmark;
Helsinki Institute of Physics (HIP), Finland;
Commissariat \`{a} l'Energie Atomique (CEA) and Institut National de Physique Nucl\'{e}aire et de Physique des Particules (IN2P3) and Centre National de la Recherche Scientifique (CNRS), France;
Bundesministerium f\"{u}r Bildung und Forschung (BMBF) and GSI Helmholtzzentrum f\"{u}r Schwerionenforschung GmbH, Germany;
General Secretariat for Research and Technology, Ministry of Education, Research and Religions, Greece;
National Research, Development and Innovation Office, Hungary;
Department of Atomic Energy Government of India (DAE), Department of Science and Technology, Government of India (DST), University Grants Commission, Government of India (UGC) and Council of Scientific and Industrial Research (CSIR), India;
National Research and Innovation Agency - BRIN, Indonesia;
Istituto Nazionale di Fisica Nucleare (INFN), Italy;
Japanese Ministry of Education, Culture, Sports, Science and Technology (MEXT) and Japan Society for the Promotion of Science (JSPS) KAKENHI, Japan;
Consejo Nacional de Ciencia (CONACYT) y Tecnolog\'{i}a, through Fondo de Cooperaci\'{o}n Internacional en Ciencia y Tecnolog\'{i}a (FONCICYT) and Direcci\'{o}n General de Asuntos del Personal Academico (DGAPA), Mexico;
Nederlandse Organisatie voor Wetenschappelijk Onderzoek (NWO), Netherlands;
The Research Council of Norway, Norway;
Commission on Science and Technology for Sustainable Development in the South (COMSATS), Pakistan;
Pontificia Universidad Cat\'{o}lica del Per\'{u}, Peru;
Ministry of Education and Science, National Science Centre and WUT ID-UB, Poland;
Korea Institute of Science and Technology Information and National Research Foundation of Korea (NRF), Republic of Korea;
Ministry of Education and Scientific Research, Institute of Atomic Physics, Ministry of Research and Innovation and Institute of Atomic Physics and University Politehnica of Bucharest, Romania;
Ministry of Education, Science, Research and Sport of the Slovak Republic, Slovakia;
National Research Foundation of South Africa, South Africa;
Swedish Research Council (VR) and Knut \& Alice Wallenberg Foundation (KAW), Sweden;
European Organization for Nuclear Research, Switzerland;
Suranaree University of Technology (SUT), National Science and Technology Development Agency (NSTDA), Thailand Science Research and Innovation (TSRI) and National Science, Research and Innovation Fund (NSRF), Thailand;
Turkish Energy, Nuclear and Mineral Research Agency (TENMAK), Turkey;
National Academy of  Sciences of Ukraine, Ukraine;
Science and Technology Facilities Council (STFC), United Kingdom;
National Science Foundation of the United States of America (NSF) and United States Department of Energy, Office of Nuclear Physics (DOE NP), United States of America.
In addition, individual groups or members have received support from:
European Research Council, Strong 2020 - Horizon 2020, Marie Sk\l{}odowska Curie (grant nos. 950692, 824093, 896850), European Union;
Academy of Finland (Center of Excellence in Quark Matter) (grant nos. 346327, 346328), Finland;
Programa de Apoyos para la Superaci\'{o}n del Personal Acad\'{e}mico, UNAM, Mexico;

\end{acknowledgement}

\bibliographystyle{utphys}   
\bibliography{bibliography}

\newpage
\appendix

\section{The ALICE Collaboration}
\label{app:collab}
\begin{flushleft} 
\small

S.~Acharya\,\orcidlink{0000-0002-9213-5329}\,$^{\rm 125}$, 
D.~Adamov\'{a}\,\orcidlink{0000-0002-0504-7428}\,$^{\rm 86}$, 
A.~Adler$^{\rm 69}$, 
G.~Aglieri Rinella\,\orcidlink{0000-0002-9611-3696}\,$^{\rm 32}$, 
M.~Agnello\,\orcidlink{0000-0002-0760-5075}\,$^{\rm 29}$, 
N.~Agrawal\,\orcidlink{0000-0003-0348-9836}\,$^{\rm 50}$, 
Z.~Ahammed\,\orcidlink{0000-0001-5241-7412}\,$^{\rm 132}$, 
S.~Ahmad\,\orcidlink{0000-0003-0497-5705}\,$^{\rm 15}$, 
S.U.~Ahn\,\orcidlink{0000-0001-8847-489X}\,$^{\rm 70}$, 
I.~Ahuja\,\orcidlink{0000-0002-4417-1392}\,$^{\rm 37}$
% ...

\section*{Affiliation Notes}

$^{\rm I}$ Deceased\\
$^{\rm II}$ Also at: Max-Planck-Institut f\"{u}r Physik, Munich, Germany\\
$^{\rm III}$ Also at: Italian National Agency for New Technologies, Energy and Sustainable Economic Development (ENEA), Bologna, Italy\\
$^{\rm IV}$ Also at: Department of Applied Physics, Aligarh Muslim University, Aligarh, India\\
$^{\rm V}$ Also at: Institute of Theoretical Physics, University of Wroclaw, Poland\\
$^{\rm VI}$ Also at: An institution covered by a cooperation agreement with CERN\\

\section*{Collaboration Institutes}

$^{1}$ A.I. Alikhanyan National Science Laboratory (Yerevan Physics Institute) Foundation, Yerevan, Armenia\\
$^{2}$ AGH University of Science and Technology, Cracow, Poland\\
$^{3}$ Bogolyubov Institute for Theoretical Physics, National Academy of Sciences of Ukraine, Kiev, Ukraine\\
$^{4}$ Bose Institute, Department of Physics  and Centre for Astroparticle Physics and Space Science (CAPSS), Kolkata, India\\
$^{5}$ California Polytechnic State University, San Luis Obispo, California, United States\\
$^{6}$ Central China Normal University, Wuhan, China\\
$^{7}$ Centro de Aplicaciones Tecnol\'{o}gicas y Desarrollo Nuclear (CEADEN), Havana, Cuba\\
$^{8}$ Centro de Investigaci\'{o}n y de Estudios Avanzados (CINVESTAV), Mexico City and M\'{e}rida, Mexico\\
$^{9}$ Chicago State University, Chicago, Illinois, United States\\
$^{10}$ China Institute of Atomic Energy, Beijing, China\\
$^{11}$ Chungbuk National University, Cheongju, Republic of Korea\\
$^{12}$ Comenius University Bratislava, Faculty of Mathematics, Physics and Informatics, Bratislava, Slovak Republic\\
$^{13}$ COMSATS University Islamabad, Islamabad, Pakistan\\
$^{14}$ Creighton University, Omaha, Nebraska, United States\\
$^{15}$ Department of Physics, Aligarh Muslim University, Aligarh, India\\
$^{16}$ Department of Physics, Pusan National University, Pusan, Republic of Korea\\
$^{17}$ Department of Physics, Sejong University, Seoul, Republic of Korea\\
$^{18}$ Department of Physics, University of California, Berkeley, California, United States\\
$^{19}$ Department of Physics, University of Oslo, Oslo, Norway\\
$^{20}$ Department of Physics and Technology, University of Bergen, Bergen, Norway\\
$^{21}$ Dipartimento di Fisica, Universit\`{a} di Pavia, Pavia, Italy\\
$^{22}$ Dipartimento di Fisica dell'Universit\`{a} and Sezione INFN, Cagliari, Italy\\
$^{23}$ Dipartimento di Fisica dell'Universit\`{a} and Sezione INFN, Trieste, Italy\\
$^{24}$ Dipartimento di Fisica dell'Universit\`{a} and Sezione INFN, Turin, Italy\\
$^{25}$ Dipartimento di Fisica e Astronomia dell'Universit\`{a} and Sezione INFN, Bologna, Italy\\
$^{26}$ Dipartimento di Fisica e Astronomia dell'Universit\`{a} and Sezione INFN, Catania, Italy\\
$^{27}$ Dipartimento di Fisica e Astronomia dell'Universit\`{a} and Sezione INFN, Padova, Italy\\
$^{28}$ Dipartimento di Fisica `E.R.~Caianiello' dell'Universit\`{a} and Gruppo Collegato INFN, Salerno, Italy\\
$^{29}$ Dipartimento DISAT del Politecnico and Sezione INFN, Turin, Italy\\
$^{30}$ Dipartimento di Scienze MIFT, Universit\`{a} di Messina, Messina, Italy\\
$^{31}$ Dipartimento Interateneo di Fisica `M.~Merlin' and Sezione INFN, Bari, Italy\\
$^{32}$ European Organization for Nuclear Research (CERN), Geneva, Switzerland\\
$^{33}$ Faculty of Electrical Engineering, Mechanical Engineering and Naval Architecture, University of Split, Split, Croatia\\
$^{34}$ Faculty of Engineering and Science, Western Norway University of Applied Sciences, Bergen, Norway\\
$^{35}$ Faculty of Nuclear Sciences and Physical Engineering, Czech Technical University in Prague, Prague, Czech Republic\\
$^{36}$ Faculty of Physics, Sofia University, Sofia, Bulgaria\\
$^{37}$ Faculty of Science, P.J.~\v{S}af\'{a}rik University, Ko\v{s}ice, Slovak Republic\\
$^{38}$ Frankfurt Institute for Advanced Studies, Johann Wolfgang Goethe-Universit\"{a}t Frankfurt, Frankfurt, Germany\\
$^{39}$ Fudan University, Shanghai, China\\
$^{40}$ Gangneung-Wonju National University, Gangneung, Republic of Korea\\
$^{41}$ Gauhati University, Department of Physics, Guwahati, India\\
$^{42}$ Helmholtz-Institut f\"{u}r Strahlen- und Kernphysik, Rheinische Friedrich-Wilhelms-Universit\"{a}t Bonn, Bonn, Germany\\
$^{43}$ Helsinki Institute of Physics (HIP), Helsinki, Finland\\
$^{44}$ High Energy Physics Group,  Universidad Aut\'{o}noma de Puebla, Puebla, Mexico\\
$^{45}$ Horia Hulubei National Institute of Physics and Nuclear Engineering, Bucharest, Romania\\
$^{46}$ Indian Institute of Technology Bombay (IIT), Mumbai, India\\
$^{47}$ Indian Institute of Technology Indore, Indore, India\\
$^{48}$ INFN, Laboratori Nazionali di Frascati, Frascati, Italy\\
$^{49}$ INFN, Sezione di Bari, Bari, Italy\\
$^{50}$ INFN, Sezione di Bologna, Bologna, Italy\\
$^{51}$ INFN, Sezione di Cagliari, Cagliari, Italy\\
$^{52}$ INFN, Sezione di Catania, Catania, Italy\\
$^{53}$ INFN, Sezione di Padova, Padova, Italy\\
$^{54}$ INFN, Sezione di Pavia, Pavia, Italy\\
$^{55}$ INFN, Sezione di Torino, Turin, Italy\\
$^{56}$ INFN, Sezione di Trieste, Trieste, Italy\\
$^{57}$ Inha University, Incheon, Republic of Korea\\
$^{58}$ Institute for Gravitational and Subatomic Physics (GRASP), Utrecht University/Nikhef, Utrecht, Netherlands\\
$^{59}$ Institute of Experimental Physics, Slovak Academy of Sciences, Ko\v{s}ice, Slovak Republic\\
$^{60}$ Institute of Physics, Homi Bhabha National Institute, Bhubaneswar, India\\
$^{61}$ Institute of Physics of the Czech Academy of Sciences, Prague, Czech Republic\\
$^{62}$ Institute of Space Science (ISS), Bucharest, Romania\\
$^{63}$ Institut f\"{u}r Kernphysik, Johann Wolfgang Goethe-Universit\"{a}t Frankfurt, Frankfurt, Germany\\
$^{64}$ Instituto de Ciencias Nucleares, Universidad Nacional Aut\'{o}noma de M\'{e}xico, Mexico City, Mexico\\
$^{65}$ Instituto de F\'{i}sica, Universidade Federal do Rio Grande do Sul (UFRGS), Porto Alegre, Brazil\\
$^{66}$ Instituto de F\'{\i}sica, Universidad Nacional Aut\'{o}noma de M\'{e}xico, Mexico City, Mexico\\
$^{67}$ iThemba LABS, National Research Foundation, Somerset West, South Africa\\
$^{68}$ Jeonbuk National University, Jeonju, Republic of Korea\\
$^{69}$ Johann-Wolfgang-Goethe Universit\"{a}t Frankfurt Institut f\"{u}r Informatik, Fachbereich Informatik und Mathematik, Frankfurt, Germany\\
$^{70}$ Korea Institute of Science and Technology Information, Daejeon, Republic of Korea\\
$^{71}$ KTO Karatay University, Konya, Turkey\\
$^{72}$ Laboratoire de Physique des 2 Infinis, Ir\`{e}ne Joliot-Curie, Orsay, France\\
$^{73}$ Laboratoire de Physique Subatomique et de Cosmologie, Universit\'{e} Grenoble-Alpes, CNRS-IN2P3, Grenoble, France\\
$^{74}$ Lawrence Berkeley National Laboratory, Berkeley, California, United States\\
$^{75}$ Lund University Department of Physics, Division of Particle Physics, Lund, Sweden\\
$^{76}$ Nagasaki Institute of Applied Science, Nagasaki, Japan\\
$^{77}$ Nara Women{'}s University (NWU), Nara, Japan\\
$^{78}$ National and Kapodistrian University of Athens, School of Science, Department of Physics , Athens, Greece\\
$^{79}$ National Centre for Nuclear Research, Warsaw, Poland\\
$^{80}$ National Institute of Science Education and Research, Homi Bhabha National Institute, Jatni, India\\
$^{81}$ National Nuclear Research Center, Baku, Azerbaijan\\
$^{82}$ National Research and Innovation Agency - BRIN, Jakarta, Indonesia\\
$^{83}$ Niels Bohr Institute, University of Copenhagen, Copenhagen, Denmark\\
$^{84}$ Nikhef, National institute for subatomic physics, Amsterdam, Netherlands\\
$^{85}$ Nuclear Physics Group, STFC Daresbury Laboratory, Daresbury, United Kingdom\\
$^{86}$ Nuclear Physics Institute of the Czech Academy of Sciences, Husinec-\v{R}e\v{z}, Czech Republic\\
$^{87}$ Oak Ridge National Laboratory, Oak Ridge, Tennessee, United States\\
$^{88}$ Ohio State University, Columbus, Ohio, United States\\
$^{89}$ Physics department, Faculty of science, University of Zagreb, Zagreb, Croatia\\
$^{90}$ Physics Department, Panjab University, Chandigarh, India\\
$^{91}$ Physics Department, University of Jammu, Jammu, India\\
$^{92}$ Physics Program and International Institute for Sustainability with Knotted Chiral Meta Matter (SKCM2), Hiroshima University, Hiroshima, Japan\\
$^{93}$ Physikalisches Institut, Eberhard-Karls-Universit\"{a}t T\"{u}bingen, T\"{u}bingen, Germany\\
$^{94}$ Physikalisches Institut, Ruprecht-Karls-Universit\"{a}t Heidelberg, Heidelberg, Germany\\
$^{95}$ Physik Department, Technische Universit\"{a}t M\"{u}nchen, Munich, Germany\\
$^{96}$ Politecnico di Bari and Sezione INFN, Bari, Italy\\
$^{97}$ Research Division and ExtreMe Matter Institute EMMI, GSI Helmholtzzentrum f\"ur Schwerionenforschung GmbH, Darmstadt, Germany\\
$^{98}$ Saga University, Saga, Japan\\
$^{99}$ Saha Institute of Nuclear Physics, Homi Bhabha National Institute, Kolkata, India\\
$^{100}$ School of Physics and Astronomy, University of Birmingham, Birmingham, United Kingdom\\
$^{101}$ Secci\'{o}n F\'{\i}sica, Departamento de Ciencias, Pontificia Universidad Cat\'{o}lica del Per\'{u}, Lima, Peru\\
$^{102}$ Stefan Meyer Institut f\"{u}r Subatomare Physik (SMI), Vienna, Austria\\
$^{103}$ SUBATECH, IMT Atlantique, Nantes Universit\'{e}, CNRS-IN2P3, Nantes, France\\
$^{104}$ Sungkyunkwan University, Suwon City, Republic of Korea\\
$^{105}$ Suranaree University of Technology, Nakhon Ratchasima, Thailand\\
$^{106}$ Technical University of Ko\v{s}ice, Ko\v{s}ice, Slovak Republic\\
$^{107}$ The Henryk Niewodniczanski Institute of Nuclear Physics, Polish Academy of Sciences, Cracow, Poland\\
$^{108}$ The University of Texas at Austin, Austin, Texas, United States\\
$^{109}$ Universidad Aut\'{o}noma de Sinaloa, Culiac\'{a}n, Mexico\\
$^{110}$ Universidade de S\~{a}o Paulo (USP), S\~{a}o Paulo, Brazil\\
$^{111}$ Universidade Estadual de Campinas (UNICAMP), Campinas, Brazil\\
$^{112}$ Universidade Federal do ABC, Santo Andre, Brazil\\
$^{113}$ University of Cape Town, Cape Town, South Africa\\
$^{114}$ University of Houston, Houston, Texas, United States\\
$^{115}$ University of Jyv\"{a}skyl\"{a}, Jyv\"{a}skyl\"{a}, Finland\\
$^{116}$ University of Kansas, Lawrence, Kansas, United States\\
$^{117}$ University of Liverpool, Liverpool, United Kingdom\\
$^{118}$ University of Science and Technology of China, Hefei, China\\
$^{119}$ University of South-Eastern Norway, Kongsberg, Norway\\
$^{120}$ University of Tennessee, Knoxville, Tennessee, United States\\
$^{121}$ University of the Witwatersrand, Johannesburg, South Africa\\
$^{122}$ University of Tokyo, Tokyo, Japan\\
$^{123}$ University of Tsukuba, Tsukuba, Japan\\
$^{124}$ University Politehnica of Bucharest, Bucharest, Romania\\
$^{125}$ Universit\'{e} Clermont Auvergne, CNRS/IN2P3, LPC, Clermont-Ferrand, France\\
$^{126}$ Universit\'{e} de Lyon, CNRS/IN2P3, Institut de Physique des 2 Infinis de Lyon, Lyon, France\\
$^{127}$ Universit\'{e} de Strasbourg, CNRS, IPHC UMR 7178, F-67000 Strasbourg, France, Strasbourg, France\\
$^{128}$ Universit\'{e} Paris-Saclay Centre d'Etudes de Saclay (CEA), IRFU, D\'{e}partment de Physique Nucl\'{e}aire (DPhN), Saclay, France\\
$^{129}$ Universit\`{a} degli Studi di Foggia, Foggia, Italy\\
$^{130}$ Universit\`{a} del Piemonte Orientale, Vercelli, Italy\\
$^{131}$ Universit\`{a} di Brescia, Brescia, Italy\\
$^{132}$ Variable Energy Cyclotron Centre, Homi Bhabha National Institute, Kolkata, India\\
$^{133}$ Warsaw University of Technology, Warsaw, Poland\\
$^{134}$ Wayne State University, Detroit, Michigan, United States\\
$^{135}$ Westf\"{a}lische Wilhelms-Universit\"{a}t M\"{u}nster, Institut f\"{u}r Kernphysik, M\"{u}nster, Germany\\
$^{136}$ Wigner Research Centre for Physics, Budapest, Hungary\\
$^{137}$ Yale University, New Haven, Connecticut, United States\\
$^{138}$ Yonsei University, Seoul, Republic of Korea\\
$^{139}$  Zentrum  f\"{u}r Technologie und Transfer (ZTT), Worms, Germany\\
$^{140}$ Affiliated with an institute covered by a cooperation agreement with CERN\\
$^{141}$ Affiliated with an international laboratory covered by a cooperation agreement with CERN.\\

\end{flushleft} 
  
\end{document}